\documentstyle[12pt,world_sci]{article}
\input{epsf}
\begin{document}

\begin{flushright}
	 NUHEP-TH-95-16 \\
	 hep-ph/9512315 \\
\end{flushright}
\bigskip
\bigskip

\title{Strange--Beauty Meson Production by \\
Fragmentation in Proton--Antiproton Collisions
\footnote{Talk presented at the International Conference on
Fronties of Physics: Looking to the 21 st Century, Shantou,
P.R. China, 5-9 August 1995.}
}
\author{
Robert J. Oakes \\
{\em Department of Physics and Astronomy, Northwestern University\\
  Evanston, IL 60208, U.S.A.\\
{\rm e-mail: oakes@fnalv.fnal.gov} }
}
\maketitle

\begin{abstract}
The cross sections for the production of $B_s$ and $B_s^*$ mesons through
the fragmentation process in $\bar{p}p$ collisions were calculated. The
input parameters were determined from the measured $B_s$ and $B_s^*$ production
rates in $e^+e^-$ annihilation. The transverse momentum distributions
$d\sigma/dp_{T}$ are given for $\sqrt{s}=1.8$ TeV, the Fermilab
Tevatron energy.
\end{abstract}
\vspace*{0.4cm}
\setlength{\baselineskip}{2.6ex}

At LEP the DELPHI Collaboration has measured the probability for a $\bar b$
antiquark to hadronize into a weakly decaying strange B meson \cite{delphi}.
Assuming the hadronization
is dominated by the fragmentation of the heavy $\bar b$ antiquark,
we determined the
fragmentation functions $D_{\bar b\to B_s}(z,\mu)$ and $D_{\bar b\to B_s^*}
(z,\mu)$ from the measured total probability for a $\bar b$ antiquark to
hadronize into the lowest strange-beauty states
$B_s(^1S_0)$ and $B_s^*( ^3S_1)$ \cite{bs}.
Here $z$ is the fraction of the $\bar b$ antiquark momentum
carried by the $B_s$ or $B_s^*$  at the scale $\mu$.
The momentum distributions of the $B_s$ and $B_s^*$ mesons produced in $e^+e^-$
annihilation at the energy of the $Z$ mass were then predicted \cite{bs}.

We then extended our calculations to the production of the $B_s$ and $B_s^*$
states in $p\bar p$ annihilation at the Tevatron~\cite{co},
using the fragmentation
functions previously determined~\cite{bs}.

In the parton model the cross section for the production by fragmentation
of strange B mesons in proton-antiproton annihilation involves
three main factors: the structure functions of the initial partons in the
proton and antiproton, the subprocess in which the initial partons produce
a particular parton in the final state, and the fragmentation of this final
parton into the strange B meson.  The transverse momentum distribution of the
$B_s$ meson is then of the form
\begin{equation}
\label{bs}
\frac{d\sigma}{dp_T} = \sum_{i,j,k} \int dx_i \,dx_j \,dz \,
f_{i/p}(x_i) f_{j/\bar p}(x_j) \frac{d\hat \sigma}{dp_T} \left(
i(x_i)j(x_j) \to k(\frac{p_T}{z})X , \mu \right) \,
D_{k\to B_s}(z;\mu) \,.
\end{equation}
Here $f_{i/p}(x_i)$ and $f_{j/\bar p}(x_j)$ are the structure functions
of the initial partons $i$ and $j$ carrying fractions of the total
momenta $x_i$ and $x_j$ in the proton and antiproton, respectively.
The production of parton $k$ with momentum $p_T/z$ is described by the
hard subprocess cross section $\hat \sigma$ at the scale $\mu$.  And
$D_{k\to B_s}(z;\mu)$ is the fragmentation function for the parton $k$ to
yield the meson $B_s$ with momentum fraction $z$ at the scale $\mu$.
The factorization scale $\mu$ is of the order of the transverse momentum
of the fragmenting parton.  The soft
physics is contained in the structure functions, while the parton subprocess
is a hard process and can be reliably calculated in perturbation theory.
We have also treated the fragmentation of $\bar b$ into $B_s$
as if it was also a hard process and perturbation theory was applicable even
though the s quark is not very heavy. The main dependence on the $s$ quark
mass is in the normalization of the fragmentation functions, which was
empirically determined~\cite{bs}.

Clearly, there is a formula very similar to Eq.~(\ref{bs}) for the production
of the $B_s^*$, which we need not write explicitly.  Because the radiative
decay $B_s^* \to B_s + \gamma$ is so fast and the $B_s^* - B_s$ mass difference
is so small,  the transverse momentum of $B_s^*$ is hardly affected in
the decay $B_s^* \to B_s +\gamma$, and thus contributes to the inclusive
production of the $B_s$.
We will, therefore, present each individually in the figures
showing $d\sigma/dp_T$ and $\sigma(p_T>p_T^{\rm min})$ and also
include their sum in a table, for convenience.

In the fragmentation process $k\to B_s$ we have included gluons as well as
$\bar b$ antiquarks.  Although the direct $g\to B_s$ fragmentation process does
not occur until order $\alpha_s^3$ there is a significant contribution
coming from the evolution of the $\bar b$ antiquark
fragmentation
functions from the heavy quark mass to the collider energy scale $Q$,
which is of order $\alpha_s^3 \log(Q/m_b)$ due to the splitting $g\to \bar b$.
This has been shown to be significant for the production of $B_c$ mesons
\cite{induce} and it was also included.

For the parton distribution functions we used the most recent CTEQ
version 3 \cite{cteq}, in which we chose the
leading-order fit, since our calculation is also a leading-order one.
For the subprocesses we used the tree-level cross sections,
to be consistent, since the parton fragmentation functions  were
calculated only to leading order.   For the
production of $\bar b$ antiquarks we included the processes
$gg\to b\bar b$, $g\bar b\to g\bar b$, and $q\bar q\to b\bar b$,
while for the production of gluons $g$ we included the processes
$gg\to gg$, $gq(\bar q) \to gq(\bar q)$, and $q\bar q\to gg$, where $q$
denotes any of the quarks, $u,d,s,c,b$.

For the running strong coupling constant $\alpha_s$ we used the simple
one-loop result evolved from its value at $\mu=m_Z$:
\begin{equation}
\alpha_s(\mu) = \frac{\alpha_s(m_Z)}{1+ \frac{33-2n_f}{6\pi} \; \alpha_s(m_Z)
\; \log \left( \frac{\mu}{m_Z} \right ) } \;.
\end{equation}
Here $n_f$ is the number of active flavors at the scale $\mu$ and we chose
$\alpha_s(m_Z)= 0.117$ \cite{pdg}.

The scale $\mu$ superficially appears to have been invented only to
separate the production of $B_s$ mesons into structure functions,
subprocess cross sections, and fragmentation functions.
In principle, this is so and the production is independent of the choice
of $\mu$.  But in practice, the results do depend on $\mu$, since  only
if all the factors are calculated to all orders in $\alpha_s$ will
the dependence on $\mu$ cancel.
However, for example, the fragmentation functions were calculated only
to leading order and, consequently,  the results will depend on $\mu$.
We investigated the sensitivity of the results to the choice of
$\mu$ by varying $\mu$ from $\mu_R/2$ to  $2 \mu_R$, where
$\mu_R=\sqrt{p_T^2({\rm parton}) +m_b^2}$ is our central choice of the scale
$\mu$.
%

To obtain the fragmentation functions at the scale $\mu$ we numerically
integrated the Altarelli-Parisi evolution equations \cite{alt} from the
scale $\mu_0$, which is of the order of the $b$-quark mass.
  As boundary conditions in these calculations we
used the fragmentation functions $D_{\bar b\rightarrow B_s}(z,\mu_0)$
and $D_{\bar b\rightarrow B_s^*}(z,\mu_0)$
at the scale $\mu_0$ calculated  using perturbative QCD~\cite{theory}.
The induced gluon fragmentation functions $D_{g\to B_s,B_s^*}(z,\mu)$ are of
order $\alpha_s^3 \log(\mu/m_b)$ and become important at large values of
the scale $\mu$ relative to the $\bar b$ antiquark fragmentation functions
$D_{\bar b\to B_s,B_s^*}(z,\mu)$, which are of order $\alpha_s^2(\mu)$.
The boundary conditions for the gluon fragmentation functions are
$
D_{g\to B_s}(z,\mu) = D_{g\to B_s^*}(z,\mu)=0
$
for $\mu \le 2(m_b+m_s)$, the threshold for producing $B_s$ or $B_s^*$ mesons
{}from a gluon.

We chose the $b$ quark mass to be $m_b=5$ GeV and the strange-quark mass
parameter $m_s$ was determined~\cite{bs} using the same
initial fragmentation functions from the experimental value \cite{delphi}
of the probability $f_s^w$ for a $\bar b$ antiquark to fragment into weakly
decaying strange-beauty mesons:
\begin{equation}
f_s^w = \int_0^1 dz \; \left[ D_{\bar b\to B_s}(z,\mu_0) +
                              D_{\bar b\to B_s^*}(z,\mu_0) \right ] \;.
\label{Y}
\end{equation}
Since the total probability for the $\bar b$ antiquark to fragment into
a $B$ meson is independent of scale, the initial scale $\mu_0$ in
Eq.~(\ref{Y}), which is of the order of the $b$-quark mass, was chosen to
be $\mu_0= m_b+2m_s$.
Then using the measured  value
$f_s^w=0.19\pm 0.06 \pm 0.08$~\cite{delphi} and
$\alpha_s(m_Z)=0.117$~\cite{pdg}
we then obtain
$
m_s = 298 +47-23 \; {\rm MeV}
$.

\begin{figure}[h]
\centering
\leavevmode
\epsfysize=240pt
\epsfbox{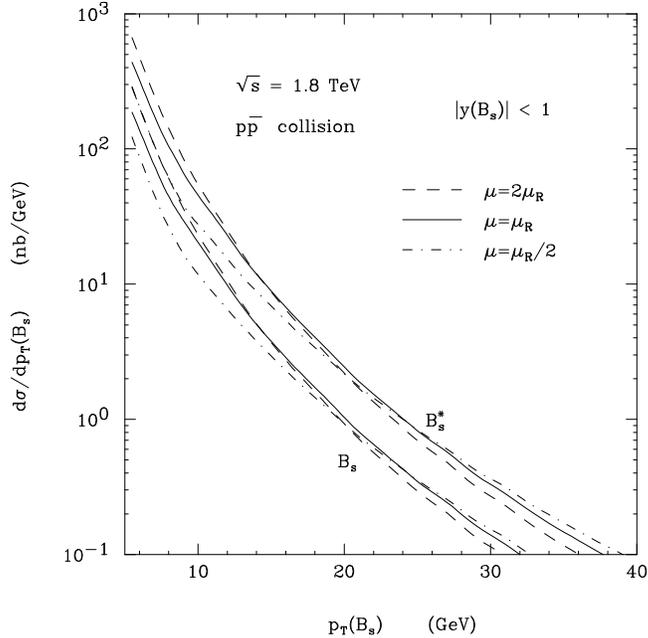}
\caption[]{
\label{fig1} \small
The transverse momentum distributions of $B_s$ and $B_s^*$ mesons
for $p_T>5$ GeV, $|y|<1$, and $\mu=\mu_R/2, \mu_R$, and $2\mu_R$ at $\sqrt{s}
=1.8$ TeV.}
\end{figure}

The evolution equations were numerically integrated and the fragmentation
functions at the scale $\mu_R=\sqrt{p_T^2({\rm parton}) + m_b^2}$ were combined
with the structure functions and parton cross sections for the subprocesses
to obtain the cross section, Eq.~(\ref{bs}).
In Fig.~\ref{fig1} we show the cross section $d\sigma/dp_T$ for both $B_s$ and
$B_s^*$.
Specifically, we calculated
the transverse momentum distributions of $B_s$ and $B_s^*$ mesons produced
in $p\bar p$ collisions
at the Tevatron energy $\sqrt{s}=1.8$~TeV.  We
assumed a cut-off on the transverse momentum of $p_T>5$ GeV$/c^2$ and
considered only the rapidity range $|y|<1$.
 To investigate the sensitivity of these results
to the scale $\mu$ we have also
included the results for $\mu=2\mu_R$ and $\mu=\mu_R/2$.
When the scale $\mu$ is less than $\mu_0=m_b+2m_s$, which only happens for the
case of $\mu=\mu_R/2$, we chose the larger of $(\mu,\mu_0)$.
{}From Fig.~\ref{fig1}
it is clear that the choice of scale $\mu$ is not critical for the
transverse momentum distribution; in fact, the variation over the range
$\mu_R/2 <\mu <2\mu_R$ is comparable to the current discrepancies between
the measured and calculated $b$ production cross sections.  As one might
expect, the $(^3S_1) \; B_s^*$ cross section  is larger than
the $(^1S_0)\;B_s$ cross section at all $p_T$,
however, their ratio is not precisely the naive prediction $3:1$,
but is about 20\% smaller.

\begin{figure}[h] 
\centering
\leavevmode
\epsfysize=260pt
\epsfbox{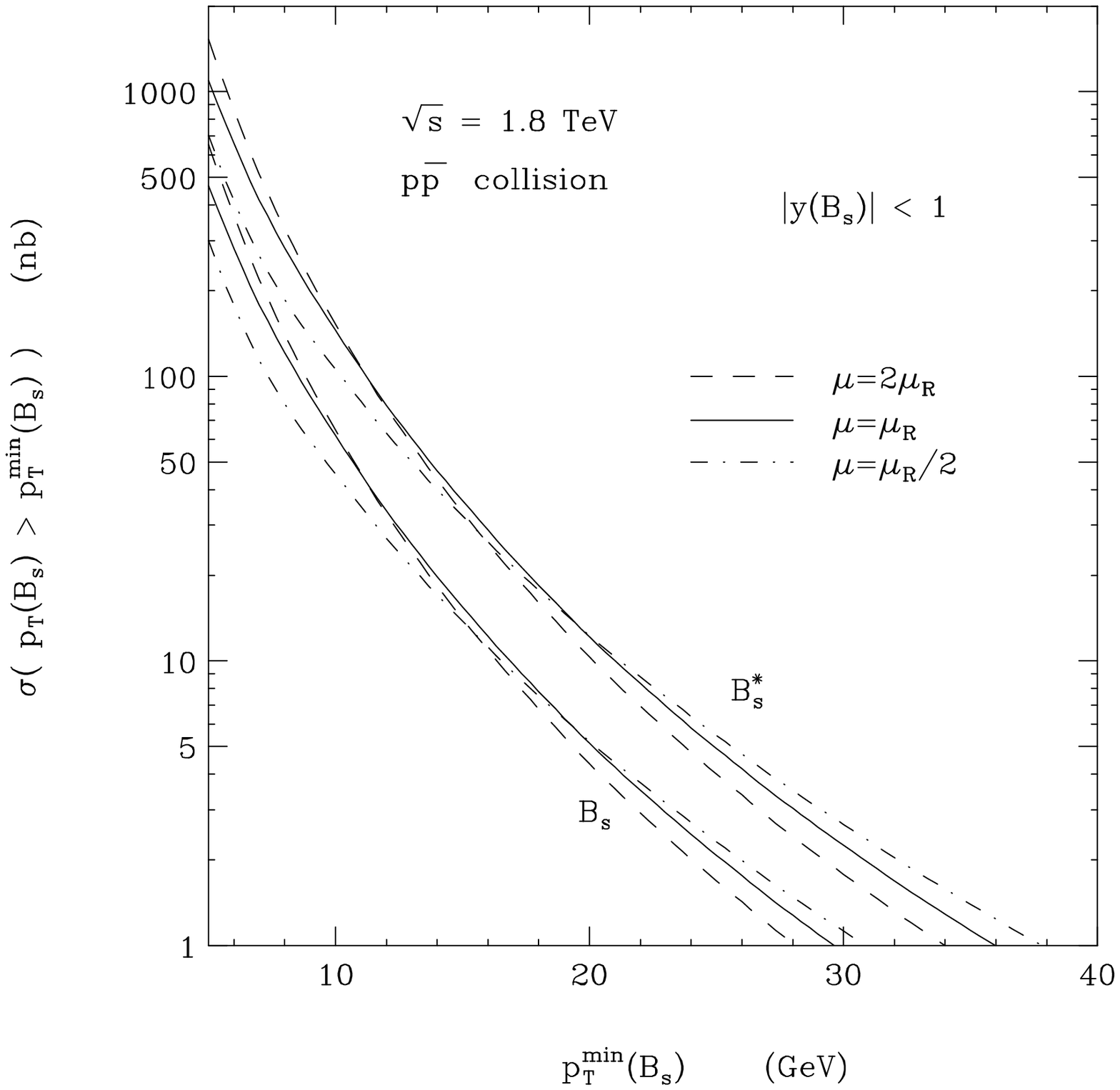}
\caption[]{
\label{fig2} \small
The $B_s$ and $B_s^*$ total cross sections for $p_T>p_T^{\rm min}$ with
$p_T>5$ GeV, $|y|<1$, and $\mu=\mu_R/2, \mu_R$, and $2\mu_R$ at $\sqrt{s}
=1.8$ TeV.}
\end{figure}

In Fig.~\ref{fig2} we show the total cross section for the production of $B_s$
and $B_s^*$ mesons with transverse momentum above a minimum value
$p_T^{\rm min}$.  As in Fig.~\ref{fig1} only the range $p_T>5$ GeV and $|y|<1$
was considered.  The sensitivity to the scale $\mu$ was investigated, as
before,
by considering $\mu=\mu_R/2$ and $\mu=2\mu_R$, and the choice of this scale
is clearly not critical.  Figure \ref{fig2} provides useful estimates
of the production rates of $B_s$ and $B_s^*$ mesons at the
Tevatron due to the fragmentation process.

\bigskip\bigskip

For convenience, we have also summarized these cross section estimates,
including the sum $\sigma(B_s) + \sigma(B_s^*)$, in Table 1.

\begin{table}[h]
\caption[]{\label{table} \small
Total $B_s$ and $B_s^*$ cross sections in nb versus $p_T^{\rm min}$ for
$p_T>5$ GeV and $|y|<1$ at $\sqrt{s}=1.8$ TeV.  Variation
with the scale $\mu$ are shown.}
\medskip
\centering
\begin{tabular}{|c|ccc|ccc|ccc|}
\hline
\hline
$p_T^{\rm min}$  &  \multicolumn{3}{c|}{$\sigma(B_s)$} & \multicolumn{3}{c|}
{$\sigma(B_s^*)$} & \multicolumn{3}{c|}{$\sigma(B_s)+\sigma(B_s^*)$} \\
\hline
  &  \underline{$\mu_R/2$} & \underline{$\mu_R$} & \underline{$2\mu_R$} &
     \underline{$\mu_R/2$} & \underline{$\mu_R$} & \underline{$2\mu_R$}  &
     \underline{$\mu_R/2$} & \underline{$\mu_R$} & \underline{$2\mu_R$}  \\
5   &   300 & 470 & 650 &    700 & 1100 & 1500  & 1000 & 1600 & 2200 \\
10  &    45 &  62 &  65 &    110 &  145 &  150  & 150  & 210  & 220 \\
15  &    14 &  16 &  14 &    32  &   37 &   33  &  46  &  52  &  47\\
20  &   5.2 &  5.1& 4.4 &    12  &   12 &   10  &  17.5&  17  &  15 \\
\hline
\end{tabular}
\end{table}

\bigskip

The production rates and transverse momentum distributions for $B_s$ and
$B_s^*$ meson production in $p\bar p$ collisions presented here assume that
fragmentation is the dominant process and, therefore, are probably
underestimates.  At
large enough transverse momentum the fragmentation process should dominate,
as it falls off more slowly, even though it is only a part of the full order
$\alpha_s^4$ contribution.
A comparative study of the relative importance of the various contributions
in the full order $\alpha_s^4$ calculation is now being carried
out~\cite{compa}, which will clarify the range of validity of the fragmentation
approximation. And the
comparison with forthcoming data from the
Fermilab Tevatron will be very instructive.

These calculations were carried out in collaboration with Kingman Cheung
and are reported in greater detailed in references [2] and [3].
This work was supported by the U.~S. Department of Energy, Division of
High Energy Physics, under Grant DE-FG02-91-ER40684 and DE-FG03-93ER40757.


\begin{thebibliography}{99}
%
\bibitem{delphi}P. Abrau {\it et al.} (DELPHI Collaboration),
Z. Phys {\bf C61}, 407 (1994).
\bibitem{bs}K. Cheung and R.J. Oakes, Phys. Lett. {\bf B337}, 181 (1994).
\bibitem{co}K. Cheung and R.J. Oakes, Phys. Rev. D, to appear.
\bibitem{induce}K. Cheung, Phys. Rev. Lett. {\bf 71}, 3413,(1993);
K. Cheung and T.C. Yuan, Phys. Lett. {\bf B325}, 481 (1994);
K. Cheung and T.C. Yuan, preprint CPP-94-37 (Feb 1995), hep-ph/9502250.
%
\bibitem{cteq}H.L. Lai {\it et al.} (CTEQ Collaboration), Phys. Rev. {\bf D51},
 4763 (1995).
\bibitem{pdg} Particle Data Group, Phys. Rev. {\bf D45}, 51 (1992).
\bibitem{alt} G. Altarelli and G. Parisi, Nucl. Phys. {\bf B126}, 298 (1977).
\bibitem{theory}E.~Braaten, K.~Cheung, and T.~C.~Yuan, Phys. Rev. {\bf D48},
 R5049 (1993).
\bibitem{compa} Y.-Q. Chen and R. J. Oakes, NUHEP-TH-95-15.
%
\end{thebibliography}
\end{document}